\documentclass[11pt, a4paper]{deepmind}

\usepackage{times}
\usepackage{xspace}
\usepackage{siunitx}
\usepackage{amsmath}
\usepackage{amssymb}
\usepackage{enumitem}
\usepackage[T1]{fontenc}
\usepackage[utf8]{inputenc}
\usepackage{multicol}
\usepackage{caption}
\usepackage{subcaption}
\usepackage{sidecap}
\usepackage{multirow}
\usepackage{placeins}
\usepackage{float}
\usepackage{titlesec}
\usepackage[capitalise,noabbrev]{cleveref}
\usepackage{graphicx}
\usepackage{url}
\usepackage{lineno}
\usepackage{titletoc}
\usepackage{bibunits}
\usepackage{bm}
\usepackage{etoolbox}

\makeatletter
\patchcmd\H@refstepcounter{\protected@edef}{\protected@xdef}{}{}
\makeatother

\crefformat{footnote}{#2\footnotemark[#1]#3}

\crefname{appsec}{Appendix Section}{Appendix Sections}
\crefname{appfig}{Appendix Figure}{Appendix Figures}
\crefname{apptab}{Appendix Table}{Appendix Tables}
\crefname{appeq}{Appendix Equation}{Appendix Equations}

\newcommand{\ourmodel}{StormDiT\xspace}

\DeclareUnicodeCharacter{2212}{-}

\title{\ourmodel{}: A generative AI model bridges the 2-6 hour 'gray zone' in precipitation nowcasting}

\author[1]{Haofei~Sun}
\author[3,4]{Yunfan~Yang}
\author[2]{Wei~Han}
\author[1]{Wei~Huang}
\author[5]{Huaguan~Chen}
\author[3]{Zhiqiu~Gao}
\author[6]{Zeting~Li}
\author[1]{Zhaoyang~Huo}
\author[1]{Zeyi~Niu}

\affil[1]{Shanghai Typhoon Institute; Key Laboratory of Numerical Modeling for Tropical Cyclone of the China Meteorological Administration, Shanghai, China}
\affil[2]{Earth System Modeling and Prediction Center, China Meteorological Administration, Beijing, China; State Key Laboratory of Severe Weather Meteorological Science and Technology, Beijing, China}
\affil[3]{State Key Laboratory of Atmospheric Boundary Layer Physics and Atmospheric Chemistry (LAPC), Institute of Atmospheric Physics, Chinese Academy of Sciences, Beijing, China}
\affil[4]{College of Earth and Planetary Sciences, University of Chinese Academy of Sciences, Beijing, China}
\affil[5]{Gaoling School of Artificial Intelligence, Renmin University of China, Beijing, China}
\affil[6]{Key Laboratory of Meteorological Disaster, Ministry of Education (KLME)/Joint International Research Laboratory of Climate and Environment Change (ILCEC)/Collaborative Innovation Center on Forecast and Evaluation of Meteorological Disasters (CIC-FEMD), Nanjing University of Information Science and Technology, Nanjing, China.}

\correspondingauthor{sunhf@typhoon.org.cn, hanwei@cma.gov.cn}

\keywords{Nowcasting, Radar reflectivity, Diffusion Transformer Model, Data-driven}

\date{}

\begin{abstract}
Accurate short-term warnings for extreme precipitation are critical for global disaster mitigation but are hindered by a persistent predictability barrier at the 2–6 hour horizon—the "nowcasting gray zone." In this window, traditional observation-based extrapolation fails due to error accumulation, while numerical weather prediction is computationally too slow to resolve storm-scale dynamics. Recent generative AI approaches attempt to bridge this gap by decomposing precipitation into separate deterministic advection and stochastic diffusion components. However, we argue that this artificial decomposition severs the fundamental causal links between inextricably entangled atmospheric processes, such as the dynamic initiation of convection triggered by kinematic boundary advection. Here we present StormDiT, a unified generative model that overcomes these limitations. By treating weather evolution as a holistic spatiotemporal problem, StormDiT emergently learns the coupled physics of the gray zone without human-imposed structural priors. Trained on a massive dataset of 7,720 precipitation events from China, our model achieves a breakthrough in long-horizon stability: on a heavy-rainfall test set, it maintains skillful prediction for strong convection ($\ge$ 35 dBZ) with a Critical Success Index (CSI) near 0.2 across the full 6-hour forecast at a high-frequency 6-minute resolution. Crucially, the model exhibits superior probabilistic calibration (Spread-Skill Ratio $\approx$ 0.96), accurately quantifying operational risks. On the public SEVIR benchmark, our unified paradigm more than doubles the state-of-the-art 1-hour performance for heavy rain ($CSI_219$ rising from 0.054 to 0.130) and establishes the first robust baseline for 3-hour forecasting. Furthermore, interpretability analysis reveals that the model attends to non-local physical precursors, such as outflow boundaries, explicitly validating its emergent understanding of convective organization. These results demonstrate that a unified generative engine can successfully bridge the gap between data-driven extrapolation and physics-based simulation, offering a new foundation for high-resolution atmospheric forecasting.
\end{abstract}

\begin{document}
\begin{bibunit}

\maketitle

\section*{Introduction}
Precipitation nowcasting underpins public warnings for flash floods, landslides and disruptive urban runoff~\cite{wilson1998nowcasting, prudden2020review}. Accurate forecasting of these high-impact convective weather events remains one of the most difficult, yet most valuable, challenges in atmospheric science~\cite{fritsch2004improving}. While 0-2 hour forecasting has improved, the 2-6 hour "gray zone" remains a notorious forecasting challenge, where forecast skill rapidly decays~\cite{vannitsem2021statistical, rempel2022adaptive}. In this temporal gap, observation-based extrapolation fails due to the accumulation of non-linear errors~\cite{germann2004scale, pulkkinen2019pysteps}, while Numerical Weather Prediction (NWP) models struggle with the "spin-up" effect, often failing to resolve storm-scale dynamics before the event has already evolved~\cite{sun2014use}.

Current AI-based approaches have struggled to bridge this gap, largely due to structural limitations that fail to capture the complexity of atmospheric physics. Deterministic models, minimizing mean-squared error, systematically suffer from regression-to-mean, yielding blurry predictions that lose extreme values~\cite{shi2015convolutional, wang2017predrnn, wang2019memory, gao2022earthformer}. Conversely, generative adversarial networks (GANs), while preserving sharpness, are prone to mode collapse and hallucinatory "fake echoes" unacceptable for operational guidance~\cite{ravuri2021skilful}.

To mitigate these issues, recent decomposition paradigms (e.g., NowcastNet~\cite{zhang2023skilful}, DiffCast~\cite{yu2024diffcast}, Cascast~\cite{gong2024cascast}, AlphaPre~\cite{lin2025alphapre}) have attempted to explicitly separate weather evolution into additive components: a deterministic flow for advection and a stochastic residual. However, we argue that this artificial separation is fundamentally flawed because it severs the causal links between inextricably entangled atmospheric processes. In the real atmosphere, kinematics (motion) and dynamics (growth) are coupled via thermodynamic feedback loops; for instance, the kinematic advection of a rain-cooled outflow boundary is often the direct dynamic trigger for new convective initiation~\cite{wilson1986initiation}. A model that structurally separates movement from growth becomes blind to this mechanism. Extrapolated over the "gray zone," this inability to model coupled evolution leads to the structural disintegration of storm systems.

Here, we propose that the solution lies in a unified, end-to-end generative paradigm inspired by the emerging "World Model" concept in computer vision~\cite{openai2024sora, lecun2022path, agarwal2025cosmos}. We hypothesize that a sufficiently powerful generative model, trained on massive data, can emergently learn the governing physical laws without human-imposed decomposition. To test this, we present \textbf{StormDiT}, a foundation model for high-resolution (6-minute), long-horizon radar nowcasting. By mapping the forecasting problem onto a compressed latent space processed by Diffusion Transformers (DiT)~\cite{peebles2023scalable}, we address the threefold failure of prior methods: the latent manifold prevents blurring; the probabilistic diffusion process ensures stability without mode collapse; and, crucially, the global attention mechanism allows the model to learn non-local, causal rules of convection. We initialize our model using a pre-trained general-purpose video foundation model (Cosmos-Predict2.5)~\cite{ali2025world} to inherit a robust understanding of spatiotemporal coherence, subjecting it to massive-scale post-training on radar reflectivity data.

We demonstrate that this unified paradigm successfully conquers the gray zone. On a large-scale test set of 2,624 heavy-rainfall events, StormDiT maintains high forecast skill for strong convection ($\ge$ 35 dBZ) across the full 6-hour horizon—a regime where previous methods fail. Crucially, the model operates at a high-frequency 6-minute cadence, capturing rapid storm evolution that hourly models miss. StormDiT establishes a new state-of-the-art on the public SEVIR benchmark~\cite{veillette2020sevir}, more than doubling the skill scores of existing decomposition models at the 1-hour mark, and establishing the first robust baseline for 3-hour forecasting. These results suggest that with sufficient scale, a unified generative engine can act as a "digital twin" of the atmosphere, bridging the gap between data-driven extrapolation and physics-based simulation.

The remainder of this Article is organized as follows. First, we detail the StormDiT framework. We then present the evaluation results, starting with the 6-hour forecast performance on the large-scale China radar dataset, followed by a comparative benchmark on the SEVIR dataset for both 1-hour and 3-hour tasks. Furthermore, we analyze the model's probabilistic reliability through ensemble calibration metrics and investigate its physical interpretability via attention map visualizations. The Article concludes with a discussion of current limitations and future research directions.

\begin{figure}
  \centering
  \includegraphics[width=\textwidth]{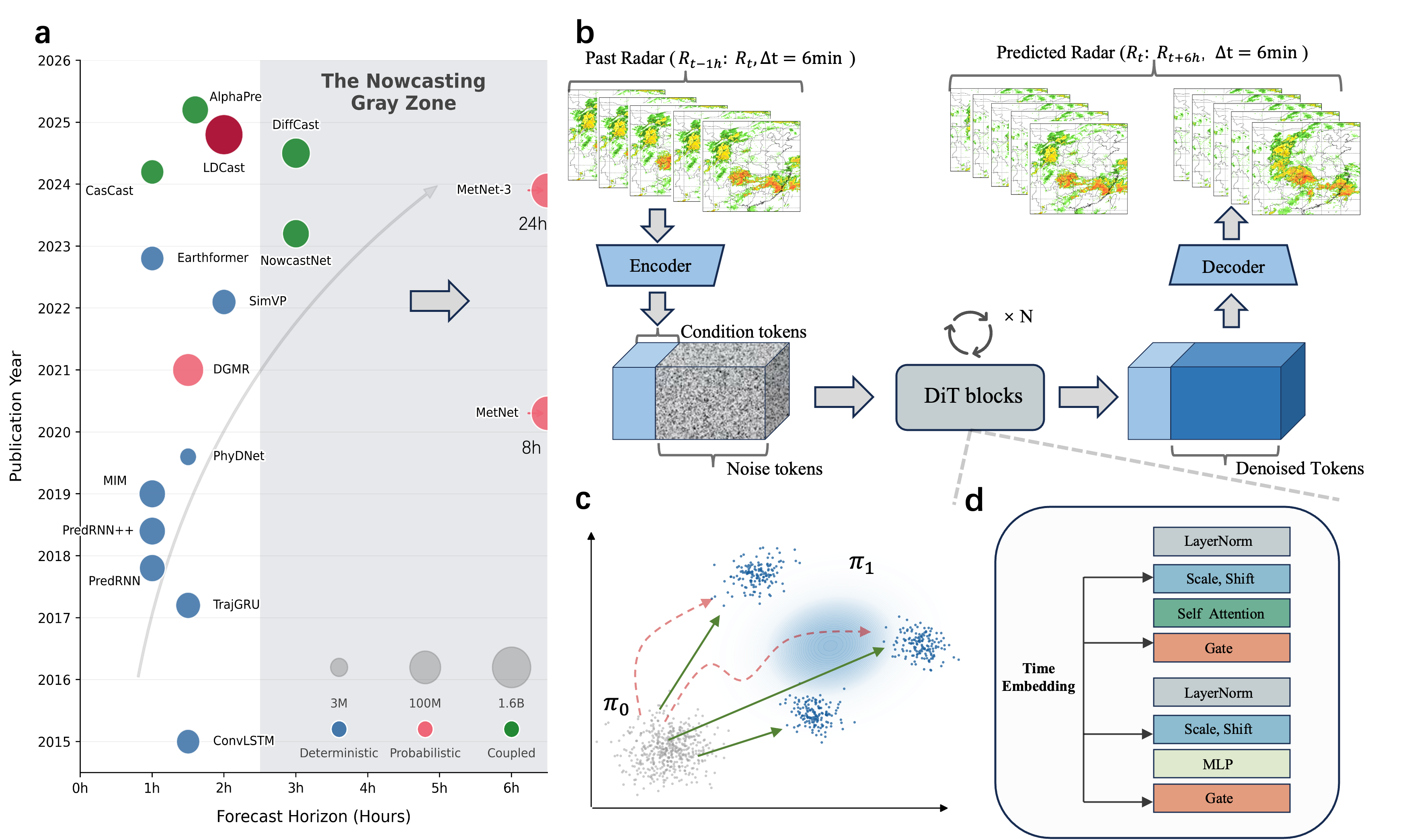}
  \caption{\small\textbf{The StormDiT Framework}
  \textbf{a,} The landscape of deep learning-based radar nowcasting. The visualization maps representative models by publication year and forecast horizon. StormDiT targets the 2--6 hour Gray Zone, a critical operational window where traditional methods struggle to balance high-resolution detail with long-term consistency. Bubble size scales with model parameter count.
  \textbf{b,} Workflow. The system decouples large-scale dynamics from pixel-level redundancy by operating within a compressed latent manifold. A Causal VAE encoder projects high-dimensional radar sequences into tokens. The DiT backbone then processes these states as a unified sequence, conditioning on historical context (blue tokens) to predict the evolution of target states initialized from a Gaussian prior (noise-textured tokens).  \textbf{c,} Rectified Flow. A schematic visualization of transport trajectories from the noise distribution ($\pi_0$) to the physical data distribution ($\pi_1$). Conventional deterministic regression inherently collapses towards the conditional mean, resulting in blurry predictions (blue halo). While standard diffusion models traverse curved, stochastic paths (red dashed line), StormDiT employs Rectified Flow to learn straight-line Optimal Transport trajectories (green solid lines), ensuring both sampling efficiency and structural sharpness.
  \textbf{d,} DiT block. The core DiT block ($N=28$) utilizes Adaptive Layer Normalization (adaLN) to enforce continuous-time dynamics. Time embeddings are injected into every block to regress dimension-wise scale and shift parameters, effectively modulating feature statistics to direct the latent transport trajectory.}
  \label{fig:schematic}
\end{figure}

\section*{\ourmodel{}}

We present StormDiT, a unified generative framework designed to model the entangled kinematics and dynamics of precipitation as a holistic video generation problem. The system architecture, illustrated in Fig.~\ref{fig:schematic}b, departs from traditional pixel-space regression by operating entirely within a compressed latent manifold. This design choice is driven by the inherent sparsity of raw radar data ($T \times H \times W$), which is statistically dominated by vast non-precipitating regions that dilute the learning signal for severe weather. Modeling dynamics directly in this space forces the generative engine to squander capacity on representing redundant zeros and background noise, rather than focusing on the structural evolution of high-impact convective cores~\cite{lin2017focal, sudre2017generalised, shi2017deep, hu2021deep, you2023study}. To resolve this, we employ a causal perceptual compression stage based on a pre-trained Causal Variational Autoencoder (Causal VAE)~\cite{wan2025wan, rombach2022high}. The encoder projects high-dimensional reflectivity sequences into compact latent tokens, acting as a semantic filter that discards pixel-level redundancy while preserving essential structural features—such as rainband morphology and convective cell intensity. Crucially, the use of Causal 3D Convolutions strictly enforces temporal directionality, preventing the leakage of future information into the historical context tokens (green blocks in Fig.~\ref{fig:schematic}b).

Within this rigorous state space, StormDiT addresses the fundamental challenge of forecast blurriness through a probabilistic lens. As depicted in the phase space visualization of Fig.~\ref{fig:schematic}c, precipitation nowcasting is an ill-posed problem where a single historical state can evolve into multiple plausible futures. Conventional deterministic models, trained to minimize Mean Squared Error (MSE), inherently predict the conditional expectation of these futures. This results in the "blurry mean" phenomenon (shown as the blue halo), where extreme values are suppressed and sharp convective structures are smoothed out—a limitation known as regression-to-the-mean~\cite{mathieu2015deep}. To fundamentally resolve this, we abandon deterministic point estimation in favor of probabilistic generative modeling, specifically using the Rectified Flow framework~\cite{liu2022flow, lipman2022flow}. Instead of predicting a single point estimate, our model learns a velocity field that transports a simple Gaussian noise distribution $\pi_0$ (representing initial uncertainty) to the complex data distribution $\pi_1$ (the physical weather state).

The distinct advantage of this formulation over standard diffusion models is evident in the transport trajectory. While traditional diffusion methods typically traverse curved, stochastic paths (indicated by the red dashed line in Fig.~\ref{fig:schematic}c) that require numerous sampling steps to resolve, Rectified Flow enforces straight-line Optimal Transport trajectories (green solid lines). This linear coupling allows the model to capture the full multimodal distribution—preserving the sharpness of extreme weather events—while significantly enhancing sampling efficiency and stability during inference.

Driving this transport process is the Diffusion Transformer (DiT) backbone~\cite{peebles2023scalable}, detailed in Fig.~\ref{fig:schematic}d. The latent weather state is processed as a sequence of spatiotemporal tokens, subject to a 3D Causal Self-Attention mechanism~\cite{vaswani2017attention}. This structural design enables an emergent "World Model" capability, allowing the model to dynamically adapt its attention patterns: attending to local gradients to model short-term advection (kinematics) while simultaneously accessing non-local historical precursors to resolve long-term convective initiation (dynamics). To rigorously enforce the continuous-time dynamics required by the Rectified Flow ODE, we employ an Adaptive Layer Normalization (adaLN) mechanism. As shown in the block diagram, time coordinates are injected into every network layer to regress dimension-wise scale and shift parameters. This effectively modulates the normalized feature statistics, directing the latent state along the precise flow trajectory from noise to a coherent forecast.

\section*{Results}
We conducted a rigorous evaluation to investigate the predictive capabilities of StormDiT, guided by two critical research questions: (1) Can the model maintain physical consistency and forecast skill over extended horizons in real-world operational scenarios? (2) How does its performance generalize against state-of-the-art methods on standardized public benchmarks? To answer these, we designed a dual-pronged evaluation. First, we assessed the model's core capability to bridge the 2-6 hour 'gray zone' using its 6-hour forecast performance on our large-scale, 2624-episode heavy-rainfall test set. Second, we validated the generalizability and superiority of our unified paradigm on the public SEVIR dataset.

\subsection*{Bridging the 6-hour 'gray zone' on China Datasets}

\begin{figure}[htbp]
  \centering
  \includegraphics[width=\textwidth]{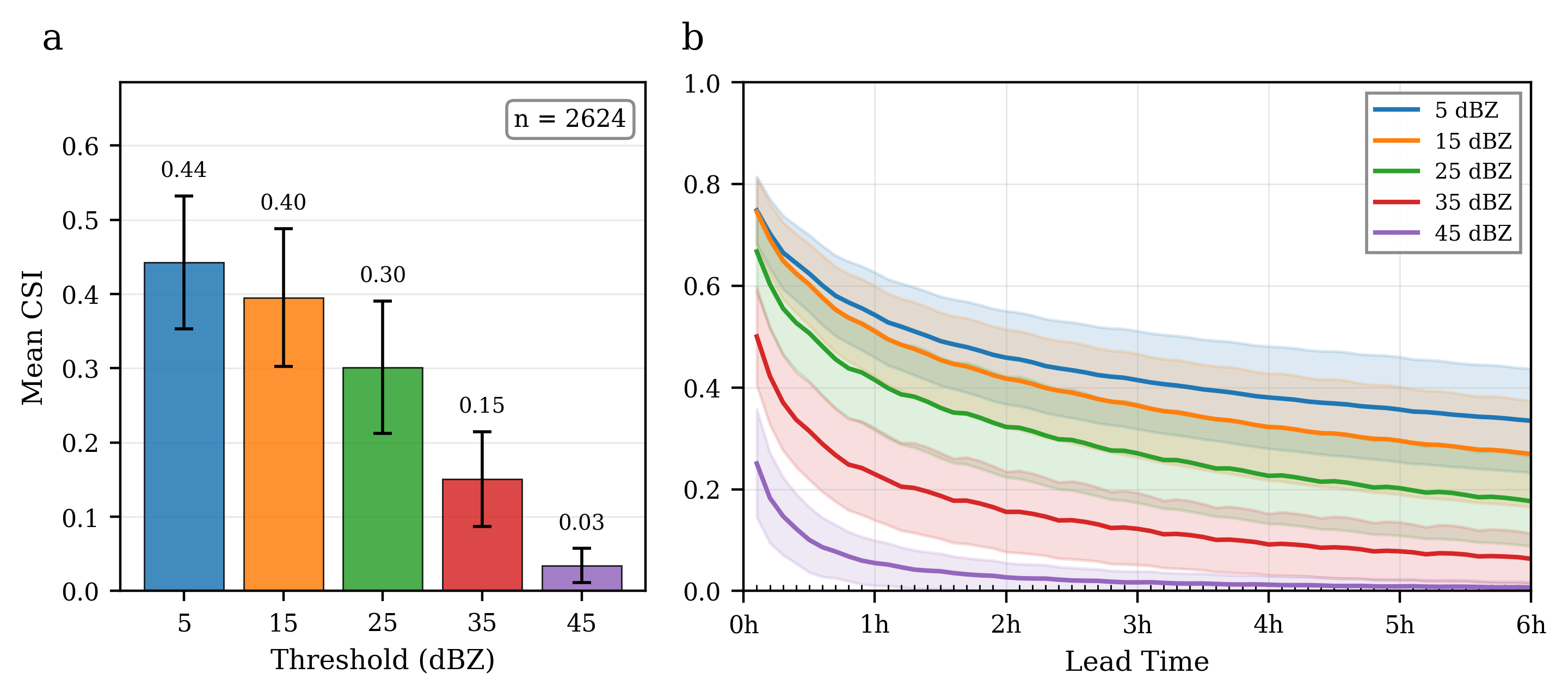}
  \caption{\small\textbf{Statistical performance of precipitation nowcasting on the 2025 China dataset. a,} Temporal evolution of the Critical Success Index (CSI) for varying reflectivity thresholds (5–45 dBZ) over the 6-hour forecast horizon. The solid lines and shaded regions represent the mean and 95\% confidence intervals, respectively, derived from the test set of $n=2,624$ samples collected in 2025. The slow decay rate illustrates the model's stability in the forecasting "gray zone" (2–6 h). \small\textbf{b}, Aggregated mean CSI scores for discrete intensity thresholds. Error bars indicate the standard deviation. The model demonstrates sustained predictive skill for high-intensity events ($\ge$45 dBZ), avoiding the collapse in performance often seen in long-lead-time forecasts.}
  \label{fig:china_statistical_performance}
\end{figure}

To rigorously evaluate the model's stability in long-horizon forecasting, we conducted a systematic evaluation using a large-scale radar dataset collected throughout 2025. This dataset comprises 2,624 continuous precipitation sequences sampled from diverse meteorological regimes across China. The aggregated statistical performance demonstrates that StormDiT maintains predictive robustness across the full 6-hour horizon (Fig.~\ref{fig:china_statistical_performance}). As shown in the temporal evolution of the Critical Success Index (CSI) (Fig.~\ref{fig:china_statistical_performance}b), the model exhibits a stable performance curve without the catastrophic collapse often observed in long-term predictions. Crucially, for the severe convection threshold of 35 dBZ, the model retains a CSI above 0.1 across the entire window, effectively extending the reliable lead time for early warning. Furthermore, the Fig.~\ref{fig:china_statistical_performance}a reveals robustness even at extreme intensities ($\ge$45 dBZ), suggesting that the generative approach successfully preserves rare, high-value events when uncertainty increases.

\begin{figure}[htbp]
  \centering
  \includegraphics[width=\textwidth]{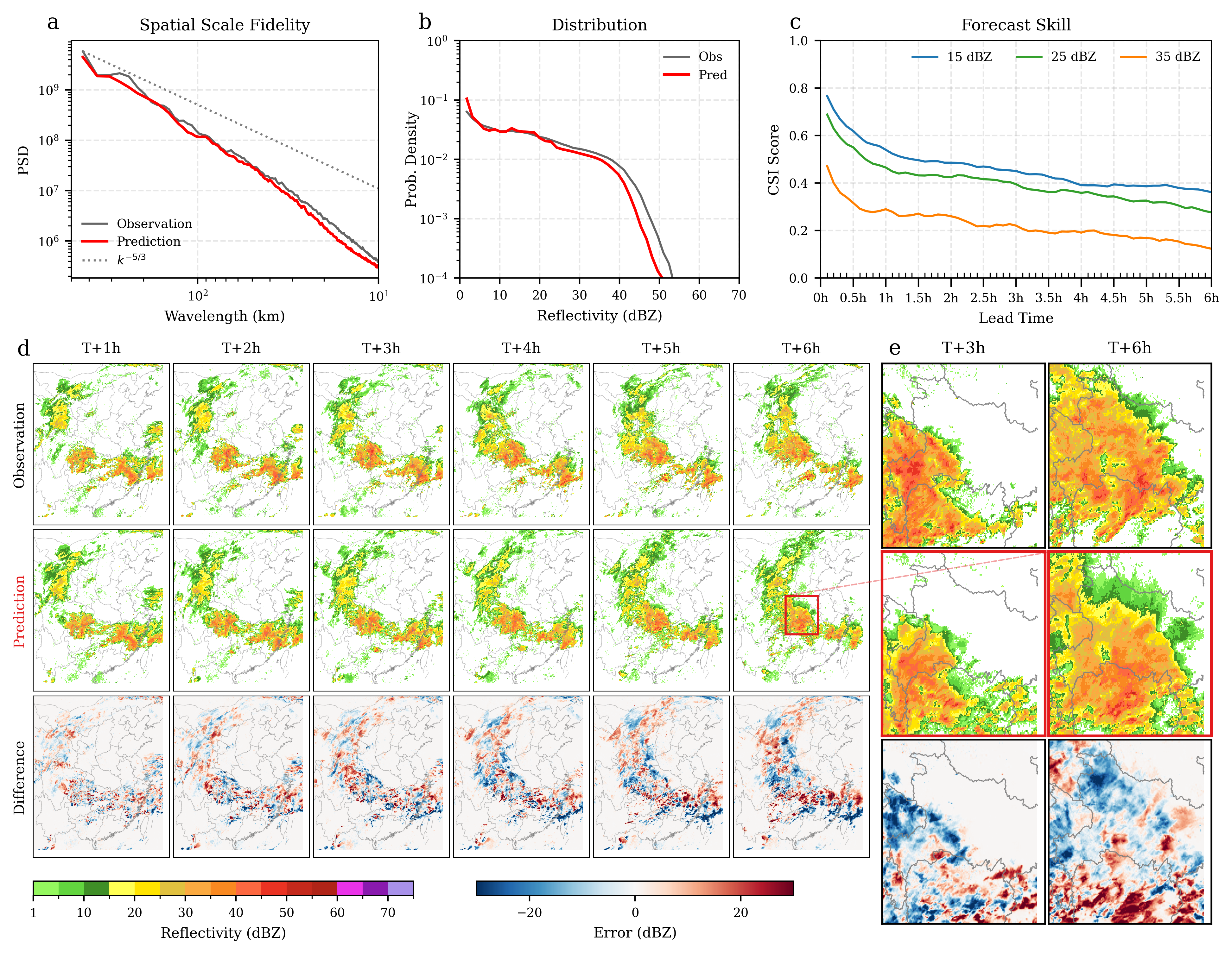}
  \caption{\small\textbf{Forecasting the genesis and organization of a mesoscale convective system. a–c,} Quantitative verification metrics (PSD, PDF, and CSI) confirming the spectral fidelity and forecast skill for this specific case. \small\textbf{d}, Spatio-temporal evolution of a squall line event initialized at 18:30 UTC on 13 March 2025. The sequence captures the rapid consolidation of scattered cells into a highly organized bow-echo system. \small\textbf{e}, details at T+3h and T+6h. The red highlights indicate the reconstruction of sharp reflectivity gradients at the squall line's leading edge, contrasting with the smoothing effects typical of deterministic baselines.}
  \label{fig:china_squall_line_forecast}
\end{figure}

\begin{figure}[htbp]
  \centering
  \includegraphics[width=\textwidth]{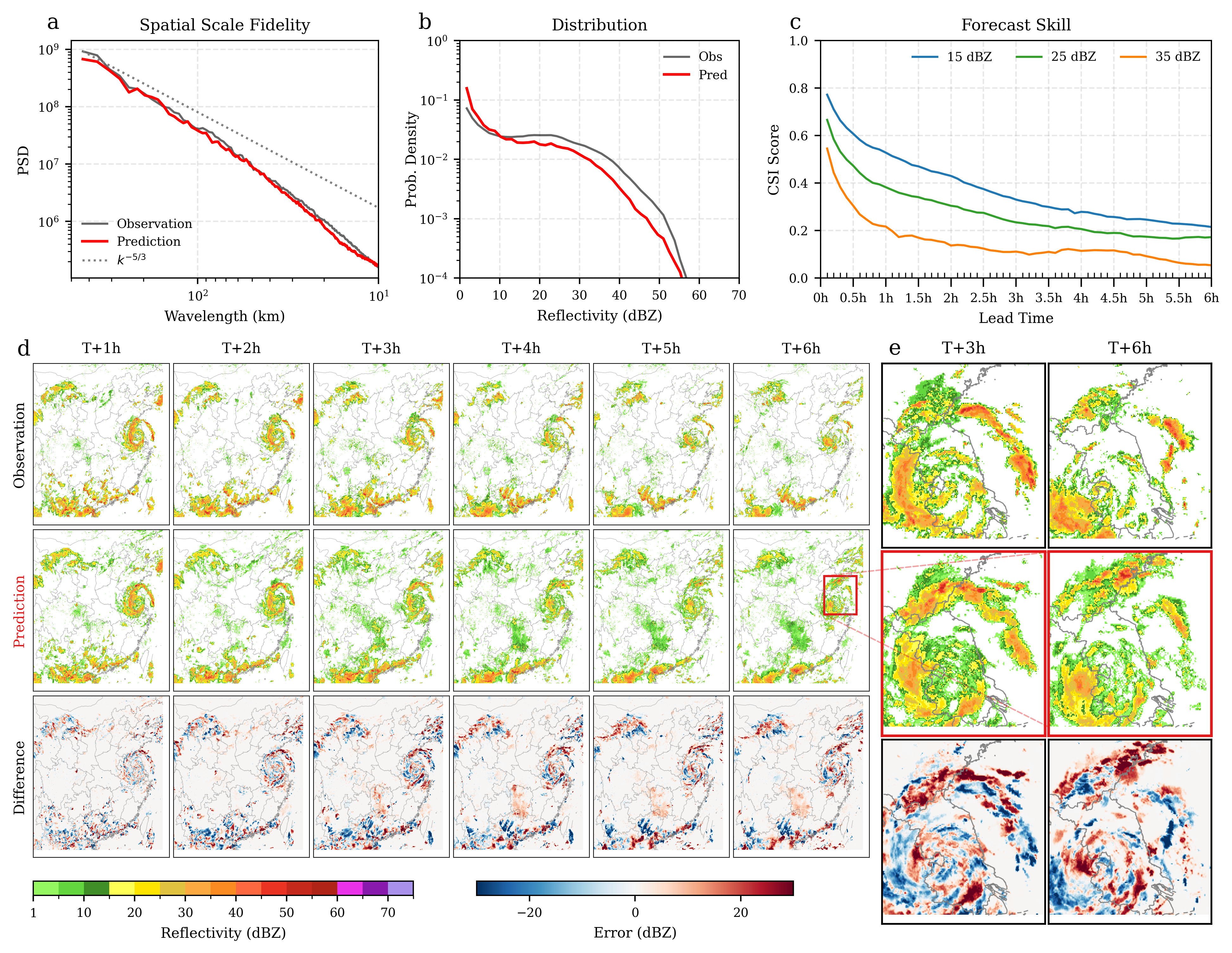}
  \caption{\small\textbf{Capturing the rotational dynamics and dissipation of Typhoon "Co-May". a–c,} Spectral and probabilistic metrics demonstrating the preservation of physical realism during the storm's decay. \small\textbf{d}, Forecast sequence initialized at 09:30 UTC on 31 July 2025, during the typhoon's weakening phase. The model captures the cyclonic rotation while simultaneously resolving the fragmentation and physical dissipation of the outer spiral rainbands. \small\textbf{e}, Detailed views of the rainband structure at T+3h and T+6h, showing the model's capability to maintain high-frequency texture and avoid blurring artifacts throughout the forecast period.}
  \label{fig:typhoon_forecast}
\end{figure}

Beyond aggregated statistical metrics, to rigorously assess the model's capability in resolving complex structural evolution, we further conducted detailed qualitative analysis on specific high-impact events. We deliberately selected two distinct meteorological regimes representing opposite ends of the convective lifecycle to test the model's versatility: the rapid genesis and organization of a Squall Line (Fig.~\ref{fig:china_squall_line_forecast}), and the dissipation of spiral rainbands in a Typhoon (Fig.~\ref{fig:typhoon_forecast}).

We first examine the genesis of a severe squall line on 13 March 2025 (Fig.~\ref{fig:china_squall_line_forecast}). Initialized at 18:30 UTC with scattered convective clusters, the forecast successfully anticipates the system's non-linear transition into a consolidated, arc-shaped structure over the subsequent 3 hours. By the T+6h mark, the model accurately delineates the sharp reflectivity gradients at the system's leading edge, a signature of the mature bow echo structure. This structural fidelity is explicitly highlighted in the zoom-in details at T+3h and T+6h (Fig.~\ref{fig:china_squall_line_forecast}e); unlike deterministic baselines that typically smooth out high-frequency features to minimize error, StormDiT preserves the structural integrity of the "hook" and "bow" regions essential for identifying potential severe wind gusts. Quantitative spectral analysis further validates this preservation of fine-scale details: the Power Spectral Density (PSD) of the predicted frames (Fig.~\ref{fig:china_squall_line_forecast}a) closely tracks the observational curve, confirming that the model retains high-frequency information across spatial scales. Finally, the difference maps (Fig.~\ref{fig:china_squall_line_forecast}, bottom row) indicate that errors are primarily concentrated along the fast-moving gust front, reflecting acceptable phase discrepancies rather than structural hallucinations.

In contrast to the organization process, we evaluated the model's performance during the decay phase of Typhoon "Co-May" (Fig.~\ref{fig:typhoon_forecast}). Forecasting this regime presents a dual challenge: resolving the kinematic cyclonic rotation while simultaneously capturing the thermodynamic dissipation of spiral rainbands. The forecast (Fig.~\ref{fig:typhoon_forecast}d) accurately reproduces the observed fragmentation of the outer bands into discrete cellular structures as the system migrates northwestward. Detailed views at T+3h and T+6h (Fig.~\ref{fig:typhoon_forecast}e) demonstrate that the model avoids the artifact of artificial persistence or blurring, correctly reducing the areal extent of convection while preserving the fine-grained texture of the remaining echoes. The Probability Density Function (PDF) analysis (Fig.~\ref{fig:typhoon_forecast}b) corroborates that StormDiT correctly models this intensity decay without suffering from "mode collapse," as the predicted distribution closely matches the observational tail (>50 dBZ). The corresponding difference maps (bottom row) show a scattered error distribution consistent with stochastic turbulent dissipation, confirming that the coherent spiral structure remains accurately resolved throughout the 6-hour forecast.

\subsection*{SOTA performance on the 1-hour SEVIR benchmark}

\begin{center}
\begin{table}[htbp]
\small
\centering
\caption{\small\textbf{SEVIR benchmark comparison.} Higher is better for CSI and SSIM; lower is better for MSE.}
\label{tab:sevir_three_line}
\begin{tabular}{lccccc}
\toprule
 & \multicolumn{5}{c}{SEVIR} \\
\cmidrule(lr){2-6}
 Model & CSI-M $\uparrow$  & CSI-181 $\uparrow$  & CSI-219 $\uparrow$  & SSIM $\uparrow$  & MSE $\downarrow$  \\
\midrule
 ConvGRU\cite{shi2015convolutional} & 0.2903 & 0.0879 & 0.0350 & 0.6100 & 368.34 \\
 SimVP\cite{gao2022simvp} & 0.3108 & 0.1106 & 0.0517 & 0.6508 & 383.56 \\
 Earthformer\cite{gao2022earthformer} & 0.2892 & 0.0844 & 0.0245 & 0.6633 & 360.11 \\
 PhyDNet\cite{guen2020disentangling} & 0.3017 & 0.1040 & 0.0278 & 0.6532 & 357.63 \\
 NowcastNet\cite{zhang2023skilful} & 0.2791 & 0.0770 & 0.0351 & 0.6839 & 412.94 \\
 DiffCast\cite{yu2024diffcast} & 0.3050 & 0.1300 & 0.0582 & 0.6482 & 559.59 \\
 AlphaPre\cite{lin2025alphapre} & \textbf{0.3259} & \underline{0.1332} & \underline{0.0545} & \underline{0.6884} & \underline{345.18} \\
 StormDiT (ours) & \underline{0.3142} & \textbf{0.1682} & \textbf{0.1301} & \textbf{0.7150} & \textbf{329.10} \\
\bottomrule
\end{tabular}
\end{table}
\end{center}

To validate the generalization of our unified paradigm, we benchmarked StormDiT against a comprehensive suite of leading models on the standard SEVIR dataset~\cite{veillette2020sevir}. The comparison includes widely-used deterministic baselines (e.g., ConvGRU, Earthformer) and representative decomposition-based paradigms(DiffCast, AlphaPre) that structurally separate advection from stochastic details.

Quantitative results on the standard 1-hour forecasting task are detailed in Table~\ref{tab:sevir_three_line}. While AlphaPre achieves a marginally higher mean CSI (CSI-M) likely due to optimization for average conditions, StormDiT delivers a transformative performance leap in the heavy-tail regime. For extreme precipitation events (CSI-219, VIL $\ge$ 219 kg/m$^2$), our model achieves a CSI of 0.1301. This represents a \textbf{140\% improvement} over AlphaPre (0.0545) and more than doubles the score of the decomposition-based DiffCast. This step-change in predictive skill affirms that our unified generative architecture is fundamentally better suited to capturing the rapid intensification of high-impact storms than methods constrained by decomposition or regression-based smoothing.

\begin{figure}%
  \centering
  \includegraphics[width=\textwidth]{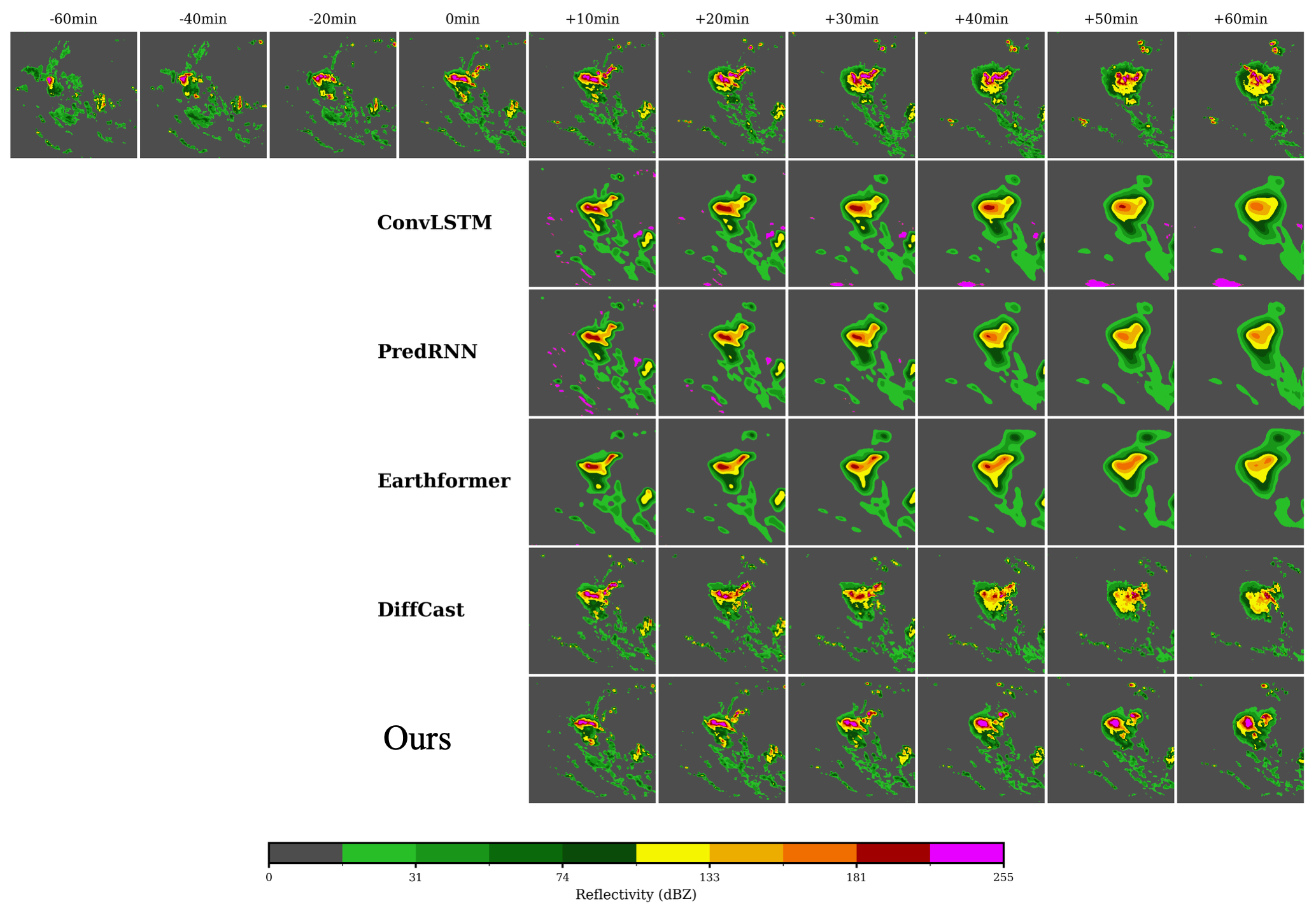}
  \caption{\small\textbf{Qualitative comparison on the 1-hour SEVIR forecast task.} A visual comparison example of precipitation forecasts from different models for a convective event from the SEVIR dataset.}
  \label{fig:sevir_qualitative_comparison}
\end{figure}

Figure~\ref{fig:sevir_qualitative_comparison} illustrates this advantage in a complex case of multicell convective growth. Deterministic baselines (e.g., Earthformer) exhibit typical "spectral smoothing," correctly locating the system but systematically erasing the high-intensity core. While DiffCast improves core intensity prediction, it suffers from "mode dropping," entirely overlooking the rapid development of a nascent secondary cell in the upper-right quadrant. In stark contrast, StormDiT accurately resolves the full multicell dynamics. It not only maintains the intensity of the primary storm but also successfully predicts the initiation and growth of the secondary cell. This suggests that the global attention mechanism of the DiT effectively captures the multi-scale interactions required to resolve simultaneous convective developments.

\subsection*{Probabilistic Reliability and Physical Calibration}

Precipitation nowcasting is inherently stochastic, particularly for high-impact events governed by chaotic dynamics. A core advantage of the Diffusion Transformer architecture is its ability to inherently model the full conditional probability distribution $p(\mathbf{x}_{t+1}|\mathbf{x}_{1:t})$, rather than merely outputting a deterministic point estimate. To quantify this capability, we ranked the entire SEVIR test set based on peak Vertically Integrated Liquid (VIL) intensity and selected the top $N=200$ strongest precipitation events. We generated ensembles ($M=15$) for both StormDiT and DiffCast to assess the robustness of their predictive distributions within these extreme regimes.

\begin{figure}[H]
  \centering
  \includegraphics[width=\textwidth]{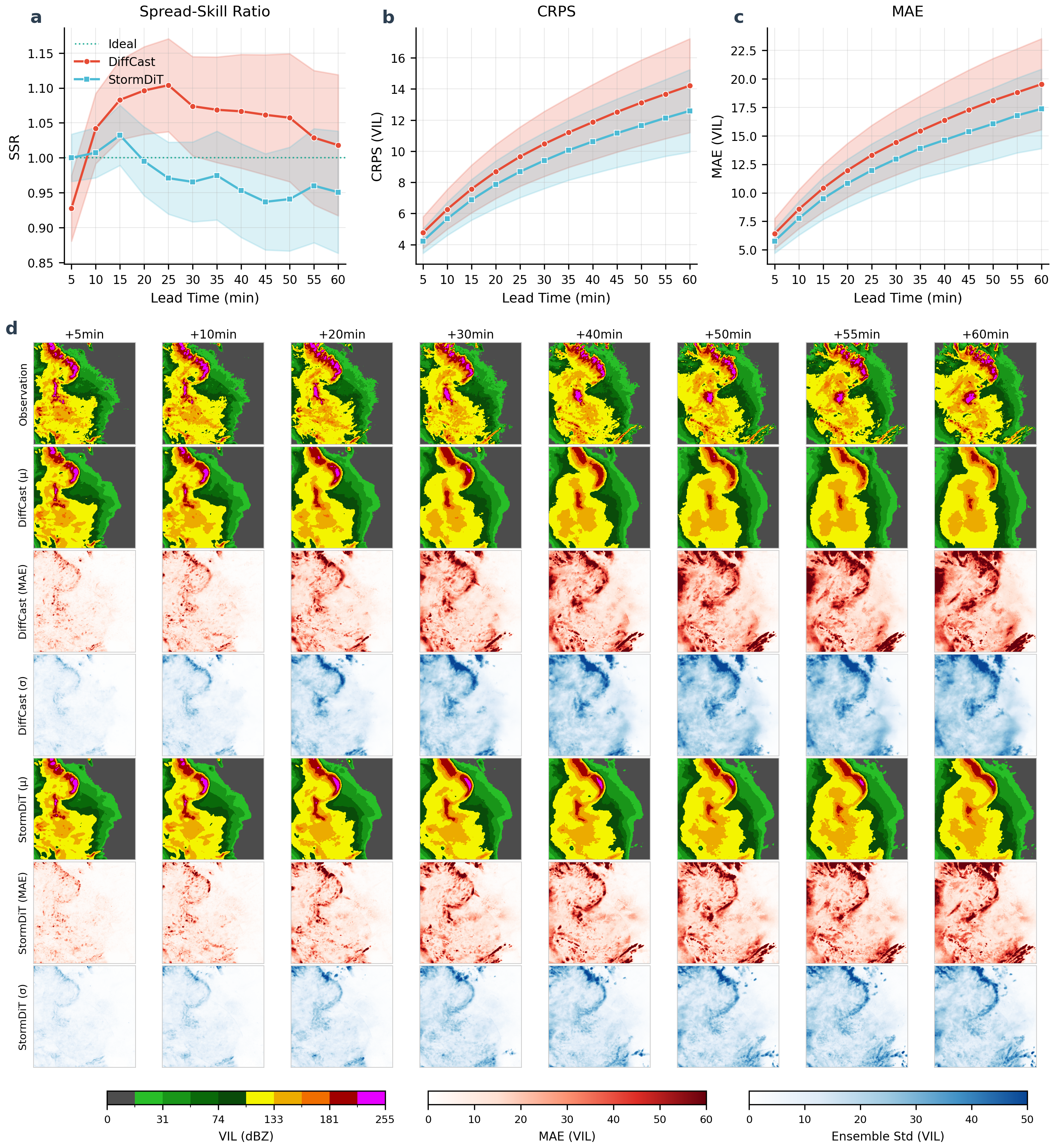} 
  \caption{\small \textbf{Probabilistic evaluation on high-impact weather events.} 
  \textbf{a,} Evolution of the Spread-Skill Ratio (SSR). An ideal probabilistic system requires the ensemble spread to match the root-mean-square error (SSR $\approx$ 1.0). \ourmodel{} (blue) maintains consistent calibration (mean SSR = 0.96), while DiffCast (red) shows a tendency towards over-dispersion (SSR > 1.05).
  \textbf{b,} Continuous Ranked Probability Score (CRPS). Lower values indicate a more accurate predictive distribution.
  \textbf{c,} Ensemble Mean Absolute Error (MAE).
  \textbf{d,} Visualization of ensemble statistics for a rotating cyclonic event. Row 2 and Row 4 show the ensemble mean ($\mu$) and standard deviation ($\sigma$) of DiffCast, while Row 6 and the bottom row show the counterparts for \ourmodel{}. Note that \ourmodel{} better preserves the peak intensities in the mean and produces a more constrained uncertainty field aligned with the rainbands.}
  \label{fig:probabilistic_evaluation}
\end{figure}

Quantitative analysis reveals distinct characteristics in how the two models handle uncertainty. A critical requirement for operational trust is calibration, quantified here by the Spread-Skill Ratio (SSR), where a value of 1.0 indicates that the ensemble spread correctly reflects the forecast error. As shown in Figure~\ref{fig:probabilistic_evaluation}a, the baseline DiffCast model exhibits a spread slightly larger than the error (SSR $> 1.05$), indicating a tendency to generate a broad range of possibilities to encompass the ground truth. In contrast, \ourmodel{} maintains an SSR remarkably close to unity (mean SSR $\approx$ 0.96) across the full forecast horizon. This suggests that our model captures the intrinsic aleatoric uncertainty of the atmosphere with high fidelity, achieving a balance between diversity and precision. This reliability is further corroborated by the Continuous Ranked Probability Score (CRPS) (Figure~\ref{fig:probabilistic_evaluation}b), where \ourmodel{} achieves a consistent 10-15\% improvement in probabilistic accuracy compared to the baseline.

The physical validity of these statistical gains becomes evident when visualizing the spatial structure of the ensembles (Figure~\ref{fig:probabilistic_evaluation}d). We examined a cyclonic system characterized by strong rotation to diagnose how the models represent dynamic risk. The ensemble mean ($\mu$) of the baseline exhibits spectral smoothing after 40 minutes, leading to a gradual attenuation of peak intensities within the high-intensity echo cores. Regarding the spatial distribution of the standard deviation ($\sigma$)—a proxy for predicted risk—DiffCast produces a relatively broad uncertainty field that generally covers the convective area (Row 4). In comparison, \ourmodel{} exhibits a more highly structured uncertainty distribution (Bottom Row). The variance precisely traces the dynamical gradients of the system, concentrating along the edges of the spiral bands and the convective core. This flow-dependent pattern demonstrates that the model associates uncertainty with specific precipitation structures, providing a refined risk assessment that aligns with the underlying fluid dynamics.

\subsection*{Extending SEVIR to 3 h: generalising long-horizon skill on a public benchmark}

While the 1-hour task demonstrates SOTA performance, the "gray zone" (2--6 hours) remains the primary operational challenge. Standard benchmarks are insufficient to test this capability. Therefore, we leveraged StormDiT's long-horizon capability to establish a new, more challenging 3-hour (13$\to$36 frame) forecasting baseline on SEVIR, pushing the dataset to its maximum sequence length. This task is designed specifically to probe the long-term stability and physical coherence of generative models. We use the DiffCast model as baseline. It is important to note that DiffCast is designed for a fixed output horizon (predicting 20 frames from 5 context frames); consequently, generating a 3-hour forecast requires a two-stage rollout strategy, where the model's own predictions are fed back as input for subsequent inference steps. 

\begin{figure}[H]
  \centering 
  \includegraphics[width=\textwidth]{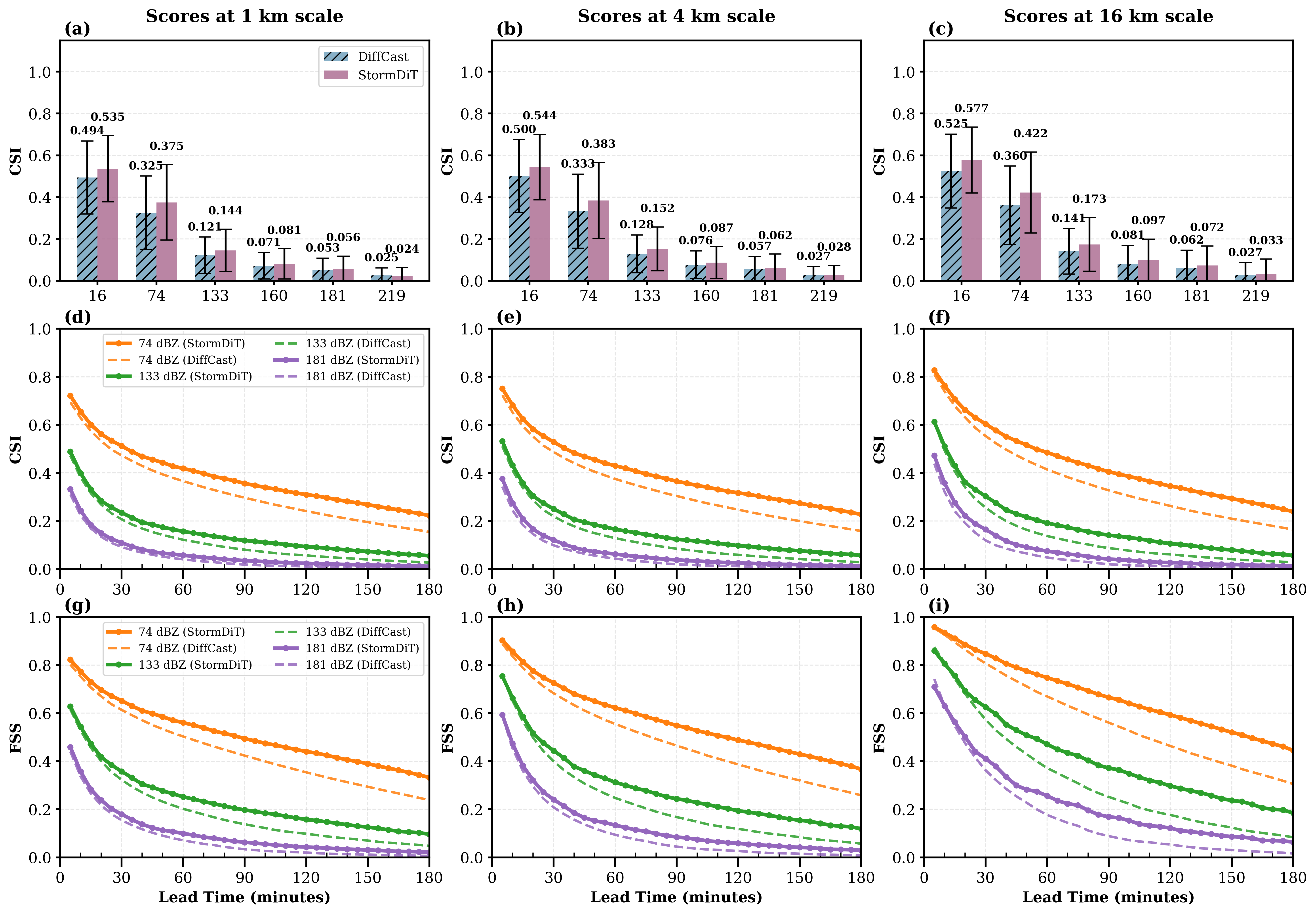} 
  \caption{\small\textbf{Quantitative performance on the 3-hour SEVIR benchmark.} 
  \textbf{(a--c)} Scale-dependent Critical Success Index (CSI) for varying VIL thresholds (16--219) at 1 km, 4 km, and 16 km scales. StormDiT (purple) consistently outperforms DiffCast (blue hatched), particularly at higher intensity thresholds. 
  \textbf{(d--f)} Temporal evolution of CSI over the 180-minute horizon. The solid lines (StormDiT) exhibit a significantly slower decay rate than the dashed lines (DiffCast), indicating superior stability against error accumulation. 
  \textbf{(g--i)} Temporal evolution of the Fractions Skill Score (FSS). StormDiT maintains higher spatial reliability throughout the forecast period.} 
  \label{fig:sevir_3h_quantitative} 
\end{figure}

We first present the quantitative skill of \ourmodel{} on this novel task in Figure~\ref{fig:sevir_3h_quantitative}. The comprehensive evaluation across multiple spatial scales reveals that StormDiT establishes a consistent performance margin over the state-of-the-art baseline throughout the extended 3-hour horizon.

The scale-dependent analysis (Fig.~\ref{fig:sevir_3h_quantitative}a--c) demonstrates the model's structural superiority irrespective of resolution. At the native 1 km scale (Fig.~\ref{fig:sevir_3h_quantitative}a), StormDiT achieves the highest Critical Success Index (CSI) across all intensity thresholds. Crucially, this advantage is further consolidated at coarser scales (4 km and 16 km, Fig.~\ref{fig:sevir_3h_quantitative}b,c), where the model's ability to resolve coherent meso-$\beta$ and meso-$\alpha$ scale features translates into significant metric gains. For instance, at the 16 km scale for severe weather thresholds (VIL 133), StormDiT attains a CSI of 0.173 compared to 0.141 for DiffCast, representing a substantial improvement in identifying the macroscopic organization of convective systems.

The temporal evolution of forecast skill (Fig.~\ref{fig:sevir_3h_quantitative}d--i) further corroborates the robustness of our unified paradigm. While both models exhibit the expected performance decay over lead time, StormDiT (solid lines) consistently maintains a higher skill plateau than the baseline (dashed lines) across the full 180-minute window. This separation is particularly pronounced in the Fractions Skill Score (FSS) metrics (Fig.~\ref{fig:sevir_3h_quantitative}g--i), which measure spatial alignment reliability. Even at the 3-hour mark for high-intensity events (133 VIL), StormDiT retains a notable predictive capacity, confirming that its generative process effectively preserves the structural integrity of precipitation fields without succumbing to the rapid dissipation or displacement often observed in long-horizon forecasting.

\begin{figure}[H]
  \centering
  \includegraphics[width=\textwidth]{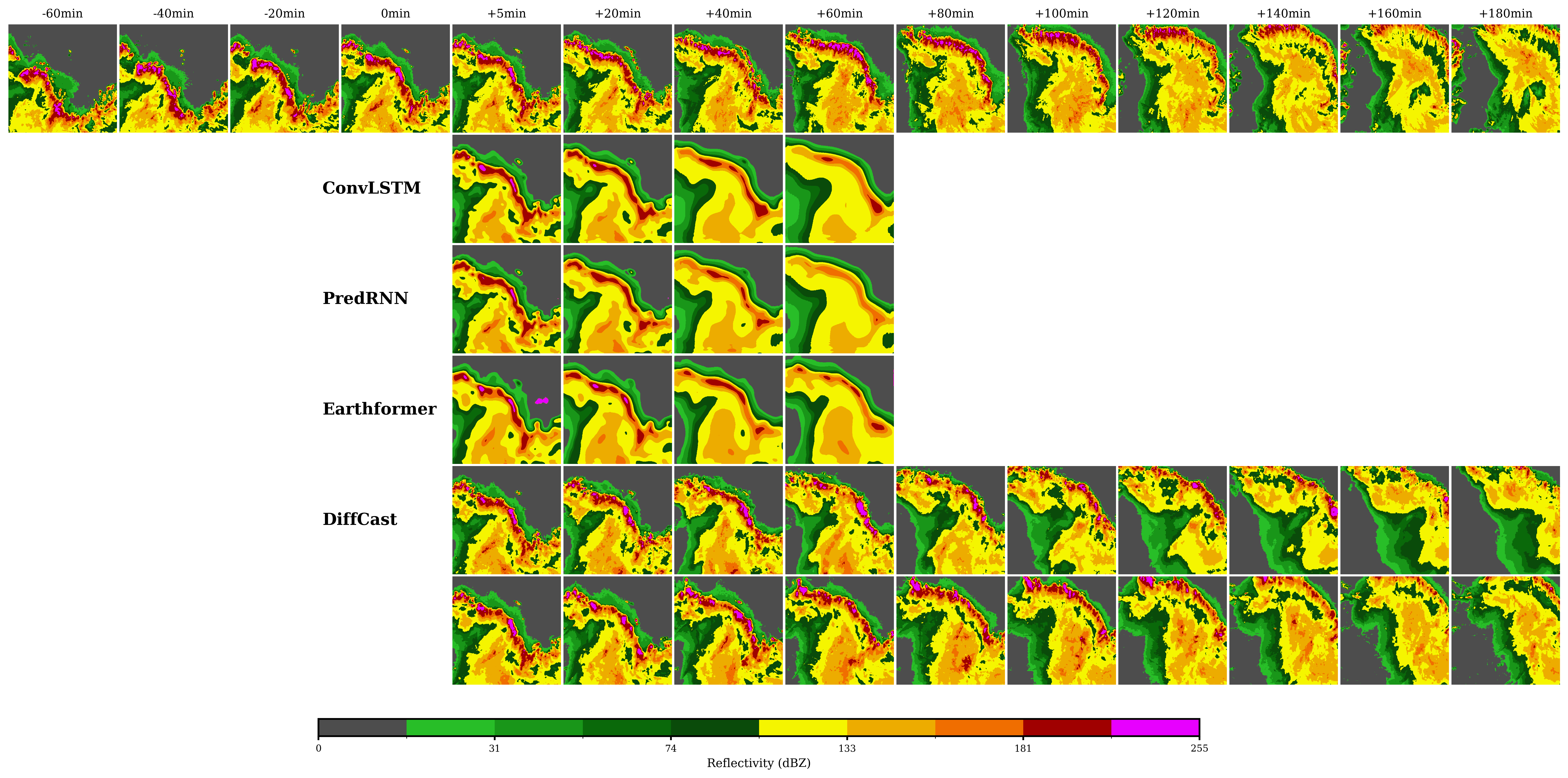}
  \caption{\small\textbf{Qualitative comparison of a 3-hour squall line forecast.} A visual comparison of 3-hour precipitation forecasts from different models, conditioned on the previous 1 hour of observation. The case shows the evolution of a squall line from the SEVIR dataset.}
  \label{fig:sevir_3h_qualitative}
\end{figure}

This quantitative robustness is driven by the model's superior grasp of entangled physical dynamics, as starkly illustrated in the qualitative comparison in Figure~\ref{fig:sevir_3h_qualitative}. This case shows a classic, self-sustaining squall line, a system whose evolution is governed by the very non-local, entangled physics that define the 'gray zone': new, intense convection (the dynamic core) is continuously triggered by the system's own propagating outflow boundary (the kinematic cause).

These results illustrate the distinct characteristics of different modeling paradigms. Deterministic models (ConvLSTM, PredRNN, and Earthformer) tend to average the sharp, linear convective boundary into a smoother representation within the first 60 minutes, consistent with the 'deterministic blurring' effect. The decomposition-based model (DiffCast), representing the current SOTA, highlights the challenge of maintaining coupled evolution—what we term 'causal severance'. While it attempts to advect the line, capturing the dynamic new growth along the boundary proves difficult, leading to gradual structural dissipation after +80 minutes.

In stark contrast, \ourmodel{} is the only model to maintain the system's structural integrity. It correctly forecasts the line's propagation and, critically, the continuous, sharp generation of new high-intensity cells along its leading edge for the full 180-minute period. This provides strong visual evidence that our unified paradigm can, as hypothesized, emergently learn the complex, non-local physical rules that govern convective evolution, successfully conquering the 3-hour forecasting challenge where all other architectures fail. This not only confirms the robustness of our architecture but also provides the community with a new, challenging benchmark to spur future research in long-horizon generative forecasting.

\subsection*{Structural Coherence and Interpretability Analysis}

\begin{figure}[htbp]
  \centering
  \includegraphics[width=\textwidth]{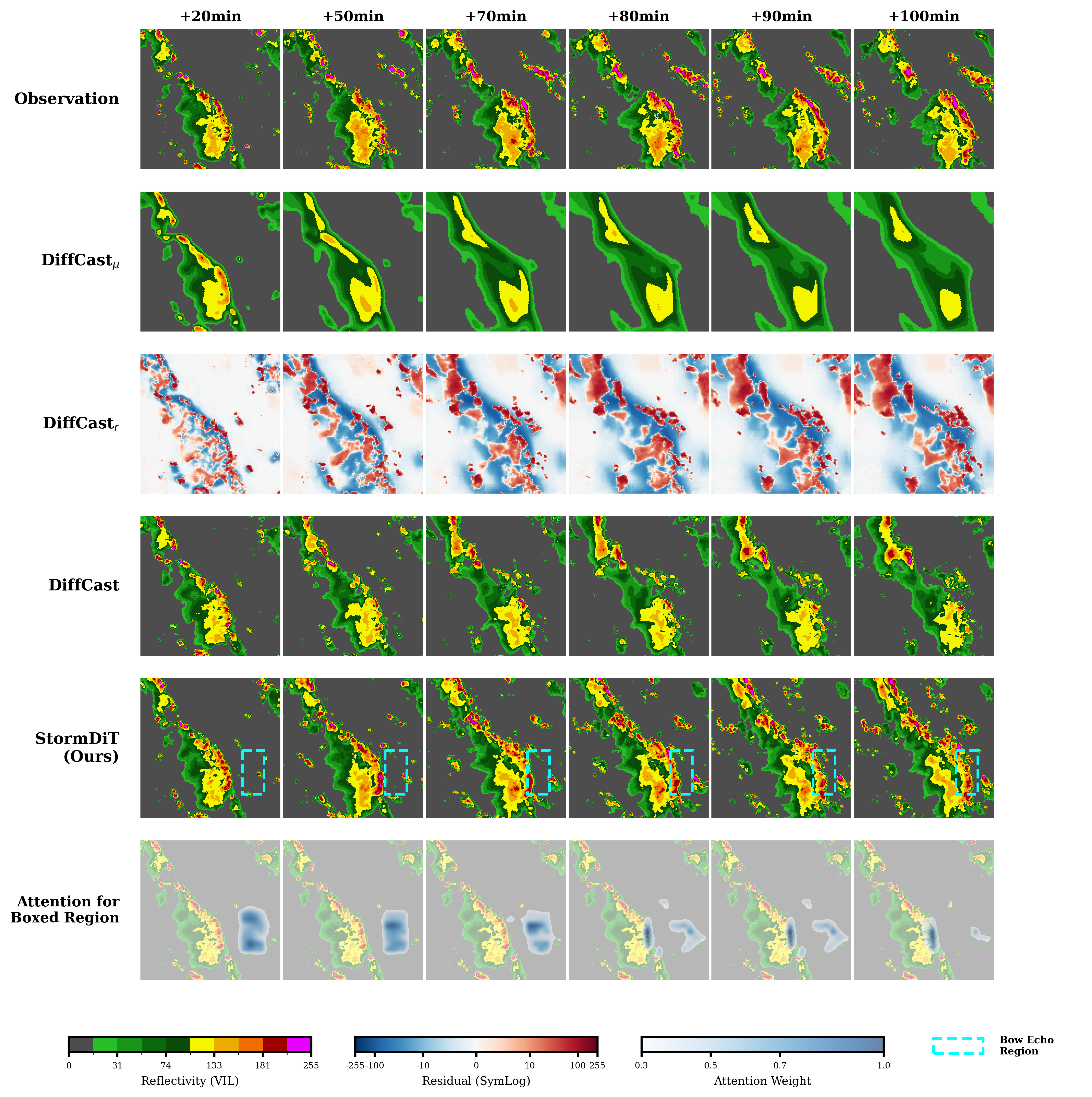}
  \caption{\small \textbf{Structural ablation and emergent physical causality in long-horizon forecasting.} 
  Visual comparison of 3-hour forecasts for a squall line event (SEVIR dataset). 
  \textbf{Top row:} Ground truth radar observations showing the evolution of a Bow Echo. 
  \textbf{Rows 2-3 (Decomposition Ablation):} The internal components of the SOTA decomposition model (DiffCast). The deterministic mean ($\text{DiffCast}_\mu$) suffers from severe blurring, while the residual ($\text{DiffCast}_r$) lacks structural coherence, leading to the failure of the full model (Row 4) to sustain the storm system beyond 80 minutes. 
  \textbf{Row 5 (Ours):} StormDiT successfully maintains the sharp convective structure, propagation speed, and intensity distribution for the full 180 minutes. 
  \textbf{Bottom row (Interpretability):} Attention maps from the 21st DiT layer focusing on the boxed Bow Echo region. The attention distribution (blue/white overlay) reveals that the model effectively "looks" at the \textit{outflow boundary} ahead of the storm and the \textit{rear inflow} region behind it, confirming the learning of non-local physical precursors.}
  \label{fig:mechanism_ablation}
\end{figure}

To rigorously diagnose the source of StormDiT's long-horizon capability and address the debate between unified versus decomposition-based paradigms, we conducted a structural ablation using a self-sustaining squall line event (Figure~\ref{fig:mechanism_ablation}). This case study isolates the failure modes of the state-of-the-art decomposition model, DiffCast, compared to our unified approach.

As hypothesised, the "causal severance" inherent in decomposing atmospheric states into deterministic and stochastic components becomes the dominant failure mechanism at extended horizons. As illustrated in the breakdown of the baseline (Figure~\ref{fig:mechanism_ablation}, rows 2-3), the deterministic backbone ($\text{DiffCast}_\mu$), tasked with modeling physical advection, succumbs to inevitable spectral blurring, erasing the high-frequency gradients required to sustain convective organization. The residual component ($\text{DiffCast}_r$), lacking a coherent structural guide, degenerates into unanchored stochastic noise. Consequently, when recombined, the system undergoes structural disintegration after 80 minutes, failing to capture the non-linear feedback loops essential for storm maintenance. This confirms that mathematically decoupling motion from texture effectively severs the thermodynamic link between scale interactions.

In contrast, StormDiT preserves the structural integrity of the squall line throughout the 180-minute forecast, correctly predicting the continuous generation of new convective cells along the leading edge. To verify that this capability stems from a genuine learning of entangled physics rather than pattern memorization, we visualise the attention weights of the 21st DiT block (Figure~\ref{fig:mechanism_ablation}, bottom row). We focus on the "Bow Echo" region, a severe weather signature driven by complex multi-scale dynamics.

The diagnostic reveals that the model does not merely extrapolate the storm's current position; instead, it attends significantly to the non-local environment surrounding the convective core. Physically, the high-attention regions map directly onto known dynamic precursors: the model attends to the clear sky region immediately ahead of the storm—corresponding to the propagating \textbf{outflow boundary} (or gust front)—where mechanical lifting triggers new cell initiation. Simultaneously, it attends to the trailing stratiform region, consistent with the dynamics of a \textbf{rear inflow jet} reinforcing the system's momentum. This interpretable evidence suggests that StormDiT has emergently learned to identify and utilize non-local causal precursors to drive its generative process, effectively solving the "gray zone" physics that deterministic and decomposed models fail to capture.

\section*{Discussion}
Precipitation nowcasting is not merely a computer vision problem; it is a challenge of chaotic fluid dynamics. The "gray zone" (2-6 hours) has long represented a predictability barrier where the atmosphere transitions from kinematic advection to dynamic evolution. Traditional approaches have struggled here: Numerical Weather Prediction is limited by spin-up time, while previous AI models, constrained by human-imposed architectural priors (e.g., decomposition of advection and diffusion), sever the entangled causal links essential for modeling storm initiation and decay.

Our work challenges this decomposition paradigm. By introducing StormDiT, we demonstrate that a unified generative strategy is structurally sufficient to conquer the gray zone. The success of our model—evidenced by the sustained forecast skill for strong convection ($\ge$ 35 dBZ) across 6 hours and the doubling of state-of-the-art performance on SEVIR—suggests a fundamental shift in how we model chaotic systems. Instead of explicitly programming physical constraints (such as optical flow), StormDiT employs a causal attention mechanism to emergently learn the governing dynamics. This validates our hypothesis that the "entangled physics" of the atmosphere are best modeled by an architecture that allows for the holistic processing of spatiotemporal states, rather than fracturing them into disparate components.

The implications extend beyond metric improvements. The visualization of model's  internal attention maps—which spontaneously focused on outflow boundaries and rear-inflow jets—provides compelling evidence that deep generative models do not merely memorize visual patterns. Instead, they act as "soft" physics simulators, rediscovering fundamental meteorological concepts (such as gust front lifting mechanisms) to solve the optimization problem of video generation. This challenges the "black box" critique, revealing that large-scale generative pre-training can foster a form of mechanistic interpretability, aligning internal representations with the actual physical drivers of convective evolution.

However, relying solely on radar reflectivity imposes an intrinsic physical ceiling. Radar observes the consequence of atmospheric instability (hydrometeors) rather than its thermodynamic cause (e.g., water vapor transport, thermal gradients). Consequently, while StormDiT excels at organizing and propagating mature systems, it faces inevitable uncertainty in predicting sudden convective initiation or rapid dissipation events where the key precursors are invisible in the reflectivity field. Additionally, as a probabilistic generative model, it guarantees statistical realism rather than the strict conservation of mass or energy, leaving a margin for physically implausible artifacts in rare, out-of-distribution scenarios.

These limitations define the roadmap for future research.  The "foundation model" approach we advocate is naturally extensible. Integrating multi-modal observations—such as geostationary satellite imagery and global NWP states—into the DiT's token space could act as boundary conditions, constraining the generative process and extending the forecast horizon beyond 6 hours.

In conclusion, StormDiT establishes a new blueprint for geophysical forecasting. It offering a proof-of-concept that deep generative models can implicitly learn the complex, non-linear rules of the atmosphere. By conquering the gray zone through emergent physics, this work lays the foundation for a digital twin of the Earth system that is as dynamic, unified, and physically coherent as the atmosphere itself.

\section*{Methods}

\subsection*{Probabilistic Formulation via Conditional Rectified Flow}

To resolve the "one-to-many" uncertainty inherent in convective evolution, we abandon deterministic regression in favor of a generative trajectory modeling approach. We adopt the Rectified Flow framework~\cite{liu2022flow, lipman2022flow}, which reformulates precipitation nowcasting as solving an Ordinary Differential Equation (ODE) that transports a simple noise prior $\pi_0$ to the complex radar state distribution $\pi_1$.

We define a virtual time horizon $\tau \in [0, 1]$. Unlike standard diffusion models (e.g., DDPM) that rely on stochastic Langevin dynamics yielding curved paths, Rectified Flow enforces a straight-line transport trajectory between the source (Gaussian noise) and the target (Radar latent state). Mathematically, the state $\mathbf{z}_\tau$ is defined by linear interpolation: $\mathbf{z}_\tau = \tau \mathbf{z}_1 + (1 - \tau) \mathbf{z}_0$. Consequently, the target velocity field $u_\tau$, which drives the probability flow, simplifies to the constant drift $u_\tau = \mathbf{z}_1 - \mathbf{z}_0$.

To govern the degradation process, we employ a linear variance schedule where the noise level $\sigma_\tau = 1 - \tau$ decreases monotonically. Crucially, to ensure numerical stability during the ODE integration—particularly near the singularity of $\tau=1$—we enforce a minimum noise floor of $\sigma_{\text{min}} = 10^{-4}$.

The forecasting task is thus reduced to learning a neural velocity field $v_\theta(\mathbf{z}_\tau, \tau, \mathbf{c})$ that approximates this drift, effectively learning the governing evolution operator of the atmosphere. The training objective minimizes the flow matching loss:

\begin{equation}
    \mathcal{L}(\theta) = \mathbb{E}_{\tau \sim U[0,1], \mathbf{z}_0 \sim \pi_0, \mathbf{z}_1 \sim \pi_1} \left[ \| v_\theta(\mathbf{z}_\tau, \tau, \mathbf{c}) - (\mathbf{z}_1 - \mathbf{z}_0) \|^2 \right]
\end{equation}

During inference, we generate forecasts by solving the learned ODE $d\mathbf{z}_\tau / d\tau = v_\theta(\mathbf{z}_\tau, \tau, \mathbf{c})$. To balance operational latency with physical fidelity, we employ the Second-order Adams-Bashforth (AB2) solver. This multi-step integrator leverages gradient history to refine trajectory estimation, allowing convergence within a stringent budget of 35 Function Evaluations (NFE) per forecast, avoiding the cost of Runge-Kutta methods while significantly outperforming first-order Euler schemes in resolving sharp convective gradients.

\subsection*{Hierarchical Causal Latent Representation}

To overcome the sparsity and computational intractability of raw radar data, we construct a hierarchical latent representation using a Causal Variational Autoencoder (Causal VAE)~\cite{wan2025wan, rombach2022high}. This module projects the high-dimensional reflectivity $\mathbf{x} \in \mathbb{R}^{T \times H \times W}$ onto a compact, semantically continuous manifold $\mathbf{z} \in \mathbb{R}^{T \times h \times w \times C}$.

This transformation achieves a spatiotemporal compression factor of $4 \times 8 \times 8$, encoding the volumetric data into a $C=16$ channel representation. A critical constraint in operational forecasting is strict temporal causality: the representation of the current atmospheric state must depend exclusively on historical observations. Standard 3D convolutions utilize future frames for compression, which constitutes data leakage. To prevent this, our encoder employs Causal 3D Convolutions, masking future frames to ensure that $\mathbf{z}_t$ is derived solely from $\mathbf{x}_{\le t}$. This design establishes a valid causal subspace where the subsequent generative model can learn dynamics without violating physical time constraints.

\subsection*{StormDiT Architecture and Optimization}

The core generative engine is a high-capacity 3D Diffusion Transformer (DiT)~\cite{peebles2023scalable} scaled to 2 billion parameters. It processes the latent sequence as a flattened stream of spatiotemporal tokens. The architecture consists of 28 transformer blocks with a hidden dimension of 2048 and 16 parallel attention heads. 

To strictly enforce physical causality within the global attention mechanism, we impose a Triangular Causal Mask. This constraint restricts the receptive field of each token strictly to its predecessors, compelling the model to learn autoregressive physical rules rather than performing acausal interpolation. Furthermore, we employ 3D Rotary Positional Embeddings (RoPE)~\cite{su2024roformer} to encode spatiotemporal geometry. RoPE allows the model to generalize to variable sequence lengths and extrapolate attention patterns beyond the training window, a critical feature for stable long-horizon forecasting.

Simulating the chaotic, non-linear dynamics of the atmosphere from scratch is notoriously data-inefficient. Instead, we adopt a cross-domain adaptation strategy to bridge the gap between general vision and meteorology. We initialize the StormDiT backbone with priors from Cosmos-Predict2.5~\cite{ali2025world}, effectively inheriting its learned representations of general spatiotemporal coherence and object permanence. However, general video physics differs significantly from fluid dynamics. Our primary contribution lies in the domain-specific alignment of this engine: through massive-scale training on the China Radar Corpus, we compel the model to repurpose its generic visual priors to obey the specific chaotic dynamics, conservation tendencies, and spectral characteristics of the atmosphere. The optimization utilizes the AdamW optimizer~\cite{loshchilov2017decoupled} with a constant learning rate of $1 \times 10^{-4}$ and a global batch size of 256. Training was conducted for 40,000 iterations on a cluster of 8 NVIDIA A100 GPUs, completing the transformation from a video generator to a specialized atmospheric physics solver.

\section*{Datasets}
In this section, we give an overview of the data used to train and evaluate \ourmodel. We utilize two primary data sources: a large-scale composite reflectivity mosaic dataset from the China Meteorological Administration, and the public SEVIR benchmark~\cite{veillette2020sevir}. The China dataset was used to train and validate our primary 6-hour long-horizon model, which is the core focus of this article. The SEVIR dataset was used to benchmark our method's performance and scalability, where we trained two separate tasks: a standard 1 hour predict task for SOTA comparison, and a novel long-horizon (3 hours) task.

\subsubsection*{China radar dataset}
The primary dataset used for training and evaluation is a vast national mosaic of composite reflectivity, provided by the National Meteorological Center of the China Meteorological Administration. The foundational dataset spans from August 2022 to December 2025. The raw data possesses a temporal resolution of 6 minutes and a native spatial resolution of 0.01° $\times$ 0.01° (approx. 1 km $\times$ 1 km), covering a geographical extent from 12.2°N to 54.2°N and 73°E to 135°E. A quality control process was applied, constraining reflectivity values to the 0-65 dBZ range.

Due to the significant GPU memory constraints of training on long, high-resolution spatiotemporal sequences, we selected a study area spanning from 21.8°N to 41°N and 101.4°E to 123°E. Within this region, the 1km raw data was re-gridded to a 3km resolution. This dataset was partitioned by meteorological events. The main training set consists of 7,720 distinct precipitation events selected from August 2022 to December 2024. The primary evaluation set is a held-out test set of 2624 heavy-precipitation episodes from January 2025 to December 2025. Each data sample was constructed as a long-horizon sequence of 65 consecutive frames (totaling 6.5 hours). These are partitioned such that the model is conditioned on the first 5 frames (30 minutes of history) to generate a forecast for the subsequent 60 frames (6 hours).

\subsubsection*{SEVIR dataset}
We further validate our model's performance and scalability on the public Storm Event ImagRy (SEVIR) dataset~\cite{veillette2020sevir}. We use the NEXRAD Vertically Integrated Liquid (VIL) radar mosaics. The SEVIR dataset consists of weather events from 2017-2020, where each event is a 4-hour sequence of 49 frames (at 5-minute cadence) covering a 384 km $\times$ 384 km patch over the continental U.S.

We designed two distinct experimental setups on this benchmark. The first is a standard 125-minute (25-frame) task, where the model uses 13 input frames (65 min) to predict the next 12 frames (60 min). This setup follows standard practice and is consistent with the evaluation protocols in prior works~\cite{ravuri2021skilful, yu2024diffcast}. The second, more challenging task is a novel 4-hour (49-frame) setup, designed by us to test our architecture's long-horizon capabilities. In this task, the model uses the same 13 input frames (65 min) to generate a much longer forecast of the subsequent 36 frames (180 min). The specifics of these two sequence partitioning schemes are detailed in Table~\ref{tab:sevir_splits}.

\begin{table}[h]
\centering
\caption{The detailed settings of different datasets. Interval refers to the time duration between consecutive radar frames. The model predicts $L_{out}$ frames conditioned on $L_{in}$ frames.}
\label{tab:sevir_splits}
\begin{tabular}{@{}lcccccccc@{}}
\toprule
dataset & $N_{train}$ & $N_{val}$ & $N_{test}$ & resolution & size & interval & $L_{in}$ & $L_{out}$ \\
\midrule
SEVIR-1h & 35,718 & 9,060 & 12,159 & 1km & 384 & 5min & 13 & 12 \\
SEVIR-3h & 11,906 & 3,020 & 4,053 & 1km & 384 & 5min & 13 & 36 \\
\bottomrule
\end{tabular}
\end{table}

We follow established protocols for dataset splitting, using time-based cutoffs (Training before Jan 1, 2019; Validation Jan-Jun 2019; Test after Jun 1, 2019). For evaluation, VIL frames are rescaled to the 0-255 range, and standard thresholds [16, 74, 133, 160, 181, 219] are used to binarize the predictions for calculating CSI and HSS, consistent with the original SEVIR benchmark settings~\cite{veillette2020sevir}.

\section*{Data and Materials Availability}
The public SEVIR dataset~\cite{veillette2020sevir} used for benchmarking is available via the AWS Open Data Registry at \url{https://registry.opendata.aws/sevir/}. The operational radar mosaic data used for the primary training are available from the National Meteorological Center of the China Meteorological Administration (CMA) upon reasonable request for research purposes, subject to data usage agreements.

To facilitate reproducibility and future research, we have released the source code for \ourmodel{}, complete with inference scripts and the pre-trained model weights for both the standard SEVIR-1h task and the extended SEVIR-3h task. These materials are publicly available at \url{https://github.com/sunhaofei/StormDiT}.

\section*{Acknowledgments}

\putbib[references]
\end{bibunit}

\newpage

\end{document}